\pdfoutput=1
\documentclass{article} 
\usepackage{amsmath,amsthm}
\usepackage{a4wide}
\usepackage{pxfonts}

\usepackage[boxruled,vlined]{algorithm2e}
\usepackage{graphicx}
\usepackage[boxruled,vlined]{algorithm2e}
\usepackage{pifont}

\newcommand{\edf}{\mbox{\textsf{EDF}}} 
\newcommand{\mote}{\mbox{\textsf{MOTE}}}

\newcommand{\equals}{\stackrel{\mathrm{def}}{=}}

\newcommand{\edff}[1]{\mbox{\textsf{EDF\(^{(#1)}\)}}}
\newcommand{\next}{\operatorname{next}}
\newcommand{\somme}{\operatorname{sum}}
\newcommand{\opt}{\operatorname{opt}}
\newcommand{\offline}{\operatorname{ol}}
\newcommand{\limit}{\operatorname{limit}}
\newcommand{\potact}{\operatorname{PotAct}}

\SetKw{Goto}{GoTo}
\SetKw{Call}{Call}
\SetKw{Continue}{Continue}
\SetKw{Et}{and}
\SetKw{Ou}{or}
\SetKw{Break}{break}

\newcommand{\busyText}[5]{
\begin{picture}(#4,3)
\put(0,0){\makebox(0,1)[c]{#1}}
\put(#4,0){\makebox(0,1)[c]{#2}}
\put(0,1){\framebox(#4,1)[c]{#5}}
\put(0,2){\makebox(#4,1){#3}}
\end{picture}
}

\newcommand{\requestText}[1]{
\begin{picture}(1,3)
\put(0,3){\vector(0,-1){1}}
\put(-1,3.5){#1}
\end{picture}
}

\newtheorem{Theorem}{Theorem} 
\newtheorem{Corollary}{Corollary}

\begin{document}

\title{Power-Aware Real-Time Scheduling\\upon Identical Multiprocessor Platforms}

\author{
\renewcommand{\thefootnote}{\arabic{footnote}}
Vincent N\'elis\footnotemark[1]~\footnotemark[2]
\and 
\renewcommand{\thefootnote}{\arabic{footnote}}
 Jo\a"el Goossens\footnotemark[1]
 \and
 \renewcommand{\thefootnote}{\arabic{footnote}}
 Nicolas Navet\footnotemark[3]
 \and
 \renewcommand{\thefootnote}{\arabic{footnote}}
 Raymond Devillers\footnotemark[1]
 \and 
 \renewcommand{\thefootnote}{\arabic{footnote}}
 Dragomir Milojevic\footnotemark[1]
}
 
\maketitle
\footnotetext[1]{Universit\'e Libre de Bruxelles (U.L.B.), CP 212, 50 Av. F. D. Roosevelt 1050 Brussels, Belgium.}
\footnotetext[2]{Supported by the Belgian National Science Foundation (FNRS) under a FRIA grant.}
\footnotetext[3]{LORIA - Equipe TRIO, Campus Scientifique - B.P. 239, 54506 Vandoeuvre-l\`es-Nancy, France.}
\addtocounter{footnote}{3}

\thispagestyle{empty}

\begin{abstract}
In this paper, we address the power-aware scheduling of sporadic constrained-deadline hard real-time tasks using dynamic voltage scaling upon multiprocessor platforms. We propose two distinct algorithms. Our first algorithm is an off-line speed determination mechanism which provides an identical speed for each processor. That speed guarantees that all deadlines are met if the jobs are scheduled using \edf. The second algorithm is an on-line and adaptive speed adjustment mechanism which reduces the energy consumption while the system is running. 
\end{abstract}

\section{Introduction}

\subsection{Context of the study}

Some important applications impose temporal constraints on the response time while running on systems with limited power resource (such as real-time communication in satellites). As a result, the research community has investigated during the past 15 years the low-power system design. Actually, the dynamic voltage scheduling (DVS) framework became a major concern for power-aware computer systems. This framework consists in minimizing the system energy consumption by adjusting the working voltage and frequency of the CPU. For real-time systems, this DVS framework focuses on minimizing the energy consumption while respecting all the timing constraints. 

Many power-constrained embedded systems are built upon multiprocessor
platforms because of high-computational requirements and because multiprocessing
often significantly simplifies the design. As pointed out in~\cite{BaAn:03}, another advantage is that multiprocessor systems
are more energy efficient than equally powerful uniprocessor
platforms, because raising the frequency of a single processor results
in a \emph{multiplicative} increase of the consumption while adding processors
leads to an \emph{additive} increase. 

\subsection{Problem definition}

In the following, we consider the problem of minimizing the energy consumption needed for executing a set of sporadic constrained-deadline real-time tasks scheduled upon a fixed number of \emph{identical} processors. The scheduling is preemptive and uses the global \edf\ policy~\cite{Liu:73}. ``Global'' scheduling algorithms, on the contrary to partitioned algorithms, allow different instances of the same task (also called jobs or processes) to be executed upon different processors. Each process can start its execution on any processor and may migrate at run-time from one processor to another if it gets meanwhile preempted by smaller-deadline processes.

We first tackle the problem of choosing the smallest (or so) processor frequency for the set of CPUs, such that all deadlines will be met. The procedure is performed off-line (i.e., before the system starts its execution) and provides a static result in the sense that the computed speed does not change over time. Such a static solution is sufficient to significantly reduce the energy consumption; however, due to the discrepancy between Worst-Case Execution Time (WCET) and Actual-Case Execution Time (ACET)~\cite{Ernst97}, it usually leads to pessimistic results. In a second step, we thus propose an on-line scheme that takes advantage of unused CPU slots to further reduce the energy consumption.

\subsection{Previous work}

There is a large number of researches about \emph{uni}processor energy-aware scheduling but much less for the multiprocessor case, where low-power scheduling problems are often NP-hard when the actual applicative constraints are taken into account~(see~\cite{ChenKuo:07} for a starting point). Among the most interesting studies, one can cite~\cite{HoonK:07} where the authors provide power-aware scheduling algorithms for bag-of-tasks applications with deadline constraints on DVS-enabled cluster systems. A study particularly relevant to the DVS framework is~\cite{ChenKuoHsu:06} which targets energy-efficient scheduling of periodic real-time tasks over multiple DVS processors with the considerations of power consumption due to leakage current (i.e. the static part of the energy dissipation). In~\cite{ChenKuo:06}, the authors propose a set of multiprocessor energy-efficient task scheduling algorithms with different task remapping and slack reclaiming schemes, where tasks have the same arrival time and share a common deadline. A large number of such {}``slack reclaiming'' approaches have been developed over the years for the \emph{uni}processor case. Among those, some strategies dynamically collect the unused computation times at the end of each job and share it among the remaining active jobs. Examples of algorithms following this {}``reclaiming'' approach, include the ones proposed in~\cite{ShCh:99,pillai01,ZhCh:02,AyMeMoMe:04}. Some reclaiming algorithms even anticipate the early completion of tasks for further reducing the CPU speed~\cite{pillai01,AyMeMoMe:04}, some having different levels of {}``aggressiveness''~\cite{AyMeMoMe:04}.

\subsection{Contribution of the paper}

Unlike the work considered in~\cite{BaAn:03}, we study the case where the number
of processors is already fixed. This constraint can be imposed by
the availability of hardware components, by design considerations
not related to power-consumption. Notice that in practical situations, the task characteristics are unknown at (hardware) design time.

The first contribution of this paper, is based on~\cite{Goossens2003Priority-driven}, and provides a technique which determines the minimum off-line processor speed for the fixed and identical multiprocessor platform using \edf.

The second, and the main contribution of this document, is a slack reclaiming algorithm which is, to the best of our knowledge, the first of its kind for the global preemptive scheduling problem of distinct-deadlines tasks on multiprocessor platforms. This contribution can be considered as an extension to the multiprocessor case of a previous proposal of Shin and Shoi in~\cite{ShCh:99}, which is usually referred to as ``One Task Extension'' (OTE). We proved that our on-line proposal does not jeopardize the system feasibility.

\paragraph{Organization of the paper.}
The document is organized as follows: in
Section~\ref{sec:model}, we introduce our model of computation, in particular our task model; 
in Section~\ref{sec:Off-line-speed-determination}, we present our off-line processor speed determination; 
in Section~\ref{sec:online}, we present our on-line speed reduction technique; 
in Section~\ref{sec:experiments}, we present our experimental results;
in Section~\ref{sec:futureworks}, we consider our future works
and in Section~\ref{sec:conclusion}, we conclude. 

\section{Model of computation}\label{sec:model}

\subsection{Application model}

We consider in this paper the scheduling of \emph{sporadic constrained-deadline tasks}, i.e., systems where each task $\tau_{i}=(C_{i}, D_{i}, T_{i})$ is characterized by three parameters -- a worst-case execution requirement (WCET) denoted $C_{i}$, a minimal inter-arrival delay $T_{i}$ and a deadline $D_{i} \leq T_i$ -- with the interpretation that the task generates successive \emph{jobs} $\tau_{i,j}$ (with $j = 1, 2, \ldots, \infty$) arriving at times $e_{i,j}$ such that $e_{i,j+1} -e_{i,j} \geq T_i$, each such job has an execution requirement of at most $C_{i}$ execution units, and must be completed by its deadline noted $D_{i,j} = e_{i,j} + D_i$. We therefore assume that the worst-case execution time is always lower than the deadline, i.e. $C_i \leq D_i$. We assume that preemption is allowed -- an executing job may be interrupted, and its execution resumed later (may be upon another processor), with no loss or penalty. Let $\tau=\{\tau_{1},\tau_{2},\ldots,\tau_{n}\}$ denotes a sporadic task system. For each task $\tau_{i}$, we define its \emph{density} $\lambda_{i}$ as the ratio of its execution requirement to its deadline: $\lambda_{i}\equals C_{i}/D_{i}$. Since $C_i \leq D_i$ we have that $\lambda_{i}\leq 1$. We also define the \emph{total density\/{}} $\lambda_{\somme}(\tau)$ of sporadic task system $\tau$ as $\lambda_{\somme}(\tau)\equals\sum_{i = 1}^{n}\lambda_{i}$, and its \emph{maximal density\/{}} as $\lambda_{\max}(\tau)\equals\max_{\tau_{i}\in\tau}\lambda_{i}$. Without loss of generality, we assume in the remainder of the paper that $\lambda_1 \geq \lambda_2 \geq \ldots \geq \lambda_n$, and consequently $\lambda_{\max}(\tau) = \lambda_1$.

\subsection{Platform model}

In our platform model, a processor can dynamically adapt its working frequency in some continuous range $\left[ f_{\min}, f_{\max} \right]$. The case where the number of frequencies is finite can be addressed as in~\cite{Navet:05}. In the remainder of this paper, we denote by $s(t)$ the processor speed at any time-instant $t$. The processor speed $s(t)$ is defined as the ratio of its current functioning frequency (say $f(t)$) over the maximal frequency $f_{\max}$, i.e.: $s(t) \equals \frac{f(t)}{f_{\max}}$, with $f_{\min} \leq f(t) \leq f_{\max}$. Notice that the processor speed always lies between $\frac{f_{\min}}{f_{\max}}$ and 1, whatever the values of $f_{\min}$ and $f_{\max}$, and to each speed corresponds exactly one frequency. 


We consider in this document multiprocessor platforms composed of a known and fixed number $m$ of identical processors $\left\{ {\cal P}_1, {\cal P}_2, \ldots, {\cal P}_m \right\}$
upon which a set of real-time tasks is scheduled. The working power of each processor may be characterized by its speed (or
computing capacity) $s$ -- with the interpretation that a job that executes on a processor of speed $s$ for $R$ time units completes $s \times R$ units of execution. The minimal and maximal admissible speed of all processors are identical and are denoted by $s_{\min} \equals \frac{f_{\min}}{f_{\max}} > 0$ and $s_{\max} \equals \frac{f_{\max}}{f_{\max}} = 1$, respectively. Since we assume that the range of available frequencies is continuous between $f_{\min}$ and $f_{\max}$, the speed of the processors can take any real value between $s_{\min}$ and $s_{\max}$ at every instant. Notice that the task computing requirements ($C_{i}$'s) are defined for the maximal speed $s_{\max}$.

In Section~\ref{sec:Off-line-speed-determination} we assume that all the processors share a common speed which is fixed before the system starts its execution. This speed does not change during the scheduling and thus, \emph{we will use the notation $s$ instead of $s(t)$ to simplify the presentation}. Then, we study the case in Section~\ref{sec:online} where each processor may run at a different speed and may change it at any time during the scheduling. In our work, speed assignments are determined at job-level: voltage/speed changes only occur at job dispatching instants. That is, once a job is assigned to a CPU, the CPU speed is fixed until the job is preempted or completed. 
\section{Off-line speed determination}
\label{sec:Off-line-speed-determination}

\subsection{Introduction}

\emph{Off-line processor speed determination} is the process of determining,
during the design of the real-time application, the lowest processor speed $s$ in order to schedule the sporadic task set $\tau$ upon an identical multiprocessor platform with $m$ processors running at speed $s$. In this Section, we consider the case where, at any instant, all processors must be running at the same speed noted $s$. We shall use the following result:

\begin{Theorem}[Bertogna, Cirinei and Lipari~\cite{ImprovedScheduleEDF}]
\label{Theorem1}
Any sporadic constrained-deadline task system $\tau$ satisfying 
\[ \lambda_{\somme}(\tau) \leq m - (m - 1) \cdot \lambda_{\max}(\tau)\]
is schedulable by the \edf\ algorithm upon a platform with $m$ identical processors.
\end{Theorem}

\noindent Then, we get the following sufficient feasibility condition:
\begin{Corollary}
\label{Corollary1}
A sporadic constrained-deadline task system $\tau$ is \edf-schedulable upon an identical multiprocessor platform with $m$ processors running at speed $s$ if:
\begin{equation}
\label{equ:Sedf}
s \geq \lambda_{\max}(\tau) + \frac{\lambda_{\somme}(\tau) - \lambda_{\max}(\tau)}{m}
\end{equation}
\end{Corollary}

Notice that, from the expression~(\ref{equ:Sedf}) (which is a sufficient condition), $s$ is always greater or equal to $\lambda_{\max}(\tau)$, which is a necessarily condition to ensure the system schedulability, whatever the scheduling algorithm. 

\subsection{Algorithm \edf$^{(k)}$}

Following an idea from~\cite{Goossens2003Priority-driven}, but
adapted to our off-line speed determination where the number of processors is \emph{fixed}, we shall present an improvement on the speed needed in order to schedule  sporadic task sets.

\paragraph{Algorithm \edff{k} (Goossens, Funk and Baruah~\cite{Goossens2003Priority-driven}):}
Assuming that the task indexes are sorted by non-increasing order of task densities and $1 \leq k \leq m$,
\edff{k} assigns priorities to jobs of tasks in $\tau$ according to the following rules: 
\begin{description}
\item [For]all $i<k$, $tau_{i}$ jobs are assigned the highest priority (ties are broken arbitrarily).
\item [For]all $i\geq k$, $\tau_{i}$ jobs are assigned priorities according to \edf\ (ties are again broken arbitrarily). 
\end{description}

That is, Algorithm~\edff{k} assigns the highest priority to jobs generated by the $(k-1)$ tasks in $\tau$ that have highest densities, and assigns priorities according to deadlines to jobs generated by all other tasks in $\tau$ (thus, {}``pure'' \edf\ is \edff{1}). We show in the following that we get another lower-bound for the speed $s$ when using \edff{k} instead of \edf, and this bound is always lower than (or equal to) the one provided by Expression~(\ref{equ:Sedf}). But first, we introduce the notation $\tau^{(i)}$ to refer to the task system composed of the $(n - i + 1)$ minimum-density tasks in $\tau$: 
$\tau^{(i)}\equals\{\tau_{i},\tau_{i+1},\ldots,\tau_{n}\}$;
(according to this notation, $\tau\equiv\tau^{(1)}$). 

\begin{Theorem}
\label{Theorem2}
Any sporadic constrained-deadline task system $\tau$ is \edf$^{(k)}$-schedulable upon an identical multiprocessor platform with $m$ processors at speed $s_k$ if $s_k \geq \max \{  \lambda_1 , \lambda_k + \frac{\lambda_{\somme}(\tau^{(k+1)})}{m - k + 1} \}$
\end{Theorem}

\begin{Corollary}
\label{cor:priD}
A sporadic constrained-deadline task system $\tau$ is schedulable upon $m$ processors at speed $s_{\offline}$ by \edff{\ell}, with
\begin{equation}
\label{equ:Soffline}\small
s_{\offline} \equals \max \{\lambda_1,\min_{k=1}^{m}\{\lambda_k + \frac{\lambda_{\somme}(\tau^{(k+1)})}{m - k + 1}\}\}
\end{equation}
and $\ell$ is the parameter minimizing the speed $s_{\offline}$ of $s_k$.
\begin{proof}
The proof is a direct consequence of Theorem~\ref{Theorem2}.
\end{proof}
\end{Corollary}

\noindent It may be seen that this expression always yields a better bound than Inequality~(\ref{equ:Sedf}).

\subsection{Implementation}
\label{sec:mrs_implementation}

A more detailed description of our off-line speed determination mechanism is given by Algorithm~\ref{ALGO_INIT}. Let $s_{\offline}$ denote the returned speed, defined by Expression~(\ref{equ:Soffline}). Before applying this algorithm, we assume that the number of processors is sufficient to schedule the system $\tau$ at the maximal speed. Consequently, the speed $s_{\offline}$ is initially set to $s_{\max}$ (line 3). Then, the algorithm searches the minimal speed by sweeping the value of $k$ between $1$ and $m$ (line 4 to line 13). Finally, in order that \edf$^{(k)}$ assigns the highest priorities to the $(k - 1)$ tasks that have highest densities, we set the deadline of these tasks to $-\infty$ (line 14).

\begin{algorithm}[h!]
\begin{footnotesize}
\linesnumbered
\KwIn{$\tau$, $m$, $s_{\max}$, $s_{\min}$}
\KwOut{$s_{\offline}$}
\Begin{
	$k_{\opt} := 1$\;
	$s_{\offline}\ := s_{\max}$ \;
	   $s_{\limit}\ := \max \{s_{\min}, \lambda_1 \}$ \;
	\For{($k := 1 $ ; $k \leq m$ \Et $s_{\offline} > s_{\limit}$ ; $k := k + 1$)}{
		$s := \max \{\lambda_1, \lambda_k + \frac{\lambda_{\somme}(\tau^{(k+1)})}{m - k + 1} \}$ \;
		\If{($s < s_{\offline}$)}{
			$s_{\offline}\ := s$ \; 
			$k_{\opt} := k$ \;
			\lIf{$(s_{\offline} < s_{\limit})$}{$s_{\offline} := s_{\limit}$ \;}
		}
	}
	\lForEach{$\tau_i \in \left\{ \tau_1,..., \tau_{k_{\opt}-1} \right\}$}{$D_i := -\infty$ \;}		
	\Return ($s_{\offline}$) \;
}
\caption{Off-line speed determination}
\label{ALGO_INIT}
\end{footnotesize}
\end{algorithm}


\section{Multiprocessor One Task Extension}
\label{sec:online}

\subsection{Introduction}

In this section, we consider the case where processors still share the same minimal and maximal speeds $s_{\min}$ and $s_{\max}$, but each one may run at its own execution speed during the scheduling. We assume that, when a processor is idle, its execution speed is always fixed to the minimal common speed $s_{\min}$. We propose a low-complexity on-line algorithm that aims to further reduce the speeds of the CPUs by performing ``local''
adjustments, when it is safe to reduce the speed below $s_{\offline}$ defined by Equation~(\ref{equ:Soffline}). 

We term our technique \mote\, for Multiprocessor One Task Extension, since it is a \emph{multiprocessor} version of the technique proposed in~\cite{ShCh:99} and usually referred to as OTE. The idea is the following: the speed of a CPU can safely be
reduced below the speed $s_{\offline}$ during the execution of a job if the reduced speed does not change anything with respect to the schedule of the subsequent jobs scheduled on that CPU. More precisely, subsequent jobs will not be delayed by more (nor less) higher-priority workload than with $s_{\offline}$. 

\subsection{Notations}

We denote by $t$ the \emph{current time} in the schedule and by $B_{i}(t)$ the \emph{last release time} of $\tau_{i}$ before or at time $t$, with $B_i(0)$ initially set to $-T_i$ (see Equation~\ref{EquationA_i} to understand this initialization). During the scheduling, $B_i(t)$ is updated at each time $t$ a job is released by $\tau_i$. The ready queue, denoted by \emph{ready-Q}, holds all the pending jobs (i.e. ready to be executed but waiting for a CPU) sorted according to the \edff{k} rule, where ties are broken according to an arbitrary rule; recall that using \edff{k}, the priorities of the jobs are \emph{constant}. In the following, $s_{i}$ denotes the processor speed for the job $\tau_{i,j}$ at time $t$. We shall use the following functions.

The function ${\cal A}_i(t, t')$ indicates \emph{if the sporadic task $\tau_{i}$ may generate a job at time $t'\geq t$}. Since $T_i$ denotes the minimal inter-arrival delay between job releases of the sporadic task $\tau_i$, we get:
\begin{equation}
\label{EquationA_i}
{\cal A}_i(t, t') \equals \left\{ \begin{array}{cl}
1 & \textrm{if}\:t' \geq B_i(t) + T_i\\
0 & \textrm{otherwise}\end{array}\right.
\end{equation}

\noindent Notice that $B_i(0)$ is initially set to $-T_i$ in order to have ${\cal A}_i(0, 0) = 1$ since our task model considers that each task may release its first job at time $t = 0$. 

Then, the function $\potact_i(t, t')$ (for \textbf{Pot}entially \textbf{Act}ive at time $t'$) indicates \emph{if $\tau_i$ has an active job at time $t$ which may still be active at time $t'$}. This function returns $1$ only if $\tau_i$ is active at time $t$ \emph{and} if $t'$ is not larger than the deadline of this job:
\[
\potact_i(t, t') \equals \left\{ \begin{array}{cl}
1 & \textrm{if}\:\omega_{i}^{s_{i}}(t) > 0\:\textrm{and}\\
& \:t \leq t' < B_i(t) + D_i \\
0 & \textrm{otherwise}\end{array}\right.
\] 
\noindent  where $\omega_{i}^{s_{i}}(t)$ denotes the \emph{remaining worst-case execution requirement} of the last released job of $\tau_i$ if executed at speed $s_i$ (if a job is done, its $\omega$ is set to zero, even if the WCET is not exhausted). 

\begin{Theorem}
\label{thm:pi}
The function 
\begin{footnotesize}
\[ \Pi(\tau_{u,v},t, t') \equals m - \sum_{\tau_i \in \tau \setminus \left\{\tau_u \right\}} \potact_i(t, t') - \sum_{\tau_i \in \tau} {\cal A}_i(t, t'), \]
\end{footnotesize}
\noindent if non-negative, provides a lower bound of the number of available CPUs at time $t'\geq t$, when ignoring the schedule of the current job of $\tau_{u}$ (if any). 
\end{Theorem}

\begin{Corollary}
\label{cor:nextnofreeprocessor}
At each time $t$ where a job $\tau_{u,v}$ is allocated to CPU ${\cal P}_{\ell}$, the earliest future time instant in the schedule such that ${\cal P}_{\ell}$ may be required by another job (possibly from the same task) is given by:
\[
t_{\next}=\left\{ \begin{array}{ll}
\min\{ t' \geq t\;|\;\Pi(\tau_{u,v},t, t') \leq 0\} & \textrm{if}\:m \leq n\\
+\infty & \textrm{otherwise}\end{array}\right.
\]
\end{Corollary}

\subsection{\mote\ scheme}
\label{sub:MOTEscheme}

\edff{k} is a job-level fixed-priority consequently a job executed on a CPU can only be preempted upon its completion or the release of a (higher priority) job. In our scheme, the speed reduction of a job is decided when the job is allocated to a CPU, for the first time or when it resumes after being preempted. Upon its release, a job is inserted into the \emph{ready-Q} if it cannot receive a processor (i.e. all processors are used and the job is of lower priority). We do not make any assumptions on the CPU allocation rule when several CPUs are available for a single job.
For instance, free CPUs can be granted according to the rule {}``smaller CPU index first.'' 

Since we consider multiprocessor platforms, we know that we have to be very careful to any change in the original schedule because of scheduling anomalies. We say that a scheduling algorithm suffers from anomalies if a change which is intuitively positive in a schedulable system can turn it unschedulable. An ``intuitively positive change'' is a change which seems to help the scheduling, like reducing the density of a task (by increasing its period or reducing its execution requirement) or advancing the start-time of a job; this can also be an increase of the number of processors on the platform. Unfortunately, multiprocessor platforms are subject to scheduling anomalies~\cite{Andersson:03}. For that reason, our on-line low-power mechanism only focuses on the last allocated-job and avoids to change the schedule of the other jobs.

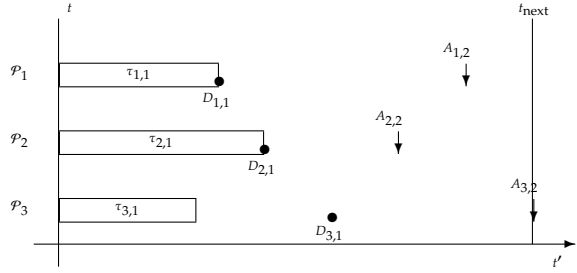
\begin{figure}[h!]
\centering
  {\tiny \setlength{\unitlength}{0.3cm}
\begin{picture}(25,10)

\put(1,0){\vector(1,0){24}}
\put(24,-1){$t'$}

\put(0,7.5){${\cal P}_1$}
\put(0,4.5){${\cal P}_2$}
\put(0,1.5){${\cal P}_3$}


\put(2.1,-1){\line(0,1){11}}
\put(2.5,10.3){$t$}

\put(2,6){\busyText{}{}{}{7}{$\tau_{1,1}$}}
\put(8.5,6.3){$D_{1,1}$}
\put(9,7){\ding{108}}
\put(20,5){\requestText{$A_{1,2}$}}

\put(2,3){\busyText{}{}{}{9}{$\tau_{2,1}$}}
\put(10.5,3.3){$D_{2,1}$}
\put(11,4){\ding{108}}
\put(17,2){\requestText{$A_{2,2}$}}

\put(2,0){\busyText{}{}{}{6}{$\tau_{3,1}$}}
\put(13.5,0.4){$D_{3,1}$}
\put(14,1){\ding{108}}
\put(23,-1){\requestText{$A_{3,2}$}}

\put(23.1,0){\line(0,1){10}}
\put(22.5,10.3){$t_{\next}$}

\end{picture}
}
\caption{Illustration of a 3-task system.}
\label{F_Principle_Example}
\end{figure}

Figure~\ref{F_Principle_Example} illustrates the main idea of our on-line algorithm when  3 tasks are scheduled upon 3 processors at speed $s_{\offline}$. This example shows a schedule where $t$ is the current time, $\tau_{1,1}$, $\tau_{2,1}$ and $\tau_{3,1}$ are the active jobs at time $t$ (the ready-queue is empty since there are only three tasks in the system) and plain circles and vertical arrows represent the deadlines and the (earliest) arrival times (since tasks are sporadic) of each task, respectively. Suppose that $\tau_{1,1}$ and $\tau_{1,2}$ are allocated to ${\cal P}_1$ and ${\cal P}_2$. Before allocating $\tau_{3,1}$ to the processor ${\cal P}_3$, we see that ${\cal P}_3$ cannot be required by another job than $\tau_{3,1}$ until time $t_{\next}$. Indeed, $\tau_{1,2}$ and $\tau_{2,2}$ could be assigned (if they arrive at time $A_{1,2}$ and $A_{2,2}$) to the CPUs ${\cal P}_1$ and ${\cal P}_2$ since the system feasibility ensures that $\tau_{1,1}$ and $\tau_{2,1}$ will be completed by their deadline. Consequently, when ignoring the schedule of $\tau_{3,1}$, we see that $t_{\next}$ is the earliest time instant (after the time $t$) such that all processors may be required. Indeed, $t_{\next}$ is the earliest time instant after time $t$ such that $\Pi(\tau_{3,1},t, t_{\next}) = 3 - 0 - 3 = 0$.

Since $t_{\next}$ is \emph{the earliest time instant} (after the current time $t$) such that ${\cal P}_3$ may be required by another job than $\tau_{3,1}$ (assuming that all the other active jobs are scheduled on other processors), one can conclude that ${\cal P}_3$ will only execute the job $\tau_{3,1}$ between time instants $t$ and $t_{\next}$. That is, we proved that ${\cal P}_3$ can modify its working speed in such a way that $\tau_{3,1}$ completes in the worst-case at time $\min\{D_{3,1}, t_{\next}\}$ (or earlier if $s_{min}$ imposes it). 

\paragraph{Principle:}
\label{sub:Principle}
Our on-line power-aware algorithm deals with a priority rule that assigns a constant priority to each job. In this work, these priorities are determined by the algorithm \edff{k}. Our power-aware algorithm is only applied when a job $\tau_{i,j}$ is to be allocated to a CPU ${\cal P}_{\ell}$ at time $t$ during the scheduling, which corresponds to its arrival or to the completion of a higher priority job. At this time, our method determines the earliest time instant $t_{\next}$ such that ${\cal P}_{\ell}$ may be needed by another job. The function $\Pi(\tau_{i,j},t, t')$ (based on the deadlines of the jobs currently executing) is used to sweep the task set (with a running  time linear in the number of tasks). Notice that the function $\Pi(\tau_{i,j},t, t')$ could be evaluated \emph{only} at the deadline-times of the jobs currently under execution and at the next (possible) arrival-time of every task (since between these instants, the function $\Pi(\tau_{i,j},t, t')$ is constant). It follows from Corollary~\ref{cor:nextnofreeprocessor} that ${\cal P}_{\ell}$ will not execute another job than $\tau_{i,j}$ until the time instant $t_{\next}$. The speed for $\tau_{i,j}$ can be safely reduced in such a way that it completes at time $\min\{D_{i,j},t_{\next}\}$ (if the corresponding speed is lower than the current one). Obviously, the working speed of a processor can never be reduced under $s_{\min}$. 

\begin{algorithm}[h!]
\begin{footnotesize}
\KwIn{$t$, $\tau_i$}
\KwOut{$t_{\next}$}
\Begin{
	$n_a := $ number of active tasks at time $t$ \;
	$L := $ set of the next deadline and possible arrival-time of each task, sorted by increasing order of the occurring time \;
	$t_{\next} := t$\;
	$\Pi := m - (n_a - 1)$\;
	\While{($\Pi > 0$ \Et $L \neq \phi$)}{
		$e \leftarrow$ L.top()\;
		$t_{\next} := $ e.occurring\_time \;
		\lIf{(e.task $\neq \tau_i$) $\Et$ (e.type == deadline)}{$\Pi := \Pi + 1$\;}
		\lElseIf{(e.type == arrival)}{$\Pi := \Pi - 1$\;}
		L.pop() \;
	}
	\Return $t_{\next}$\;
}
\end{footnotesize}
\caption{Determination of $t_{\next}$}
\label{ALGO_Tnext}
\end{algorithm}

\begin{algorithm}[h!]
\begin{footnotesize}
\KwIn{$\tau_{i,j}$}
\KwOut{$\phi$}
\Begin{
	// Initialization step \\ 
	\If{($\tau_{i,j}$ is allocated for the first time)}{
		\lIf{($i<k$)}{$s_i := \lambda_i$\;}
		\lElse{$s_i := \lambda_k + \frac{\lambda_{\somme}(\tau^{(k+1)})}{m - k + 1}$ \;}
	}
	// \mote\ step \\
	   \lIf{($m \leq n$)}{$t_{\next} := $ \Call Algorithm2($t$, $\tau_i$)} \;
	\lElse{$t_{\next} := \infty$ \;}
	 \If{($t_{\next} > t$)}{
		$s_{i} := \min\{s_i,\frac{\omega_{i}^{s_{i}}(t) \cdot s_{i}}{\min\{D_{i,j},t_{\next}\}-t} \}$ \;
		\lIf{($s_i < s_{\min}$)}{$s_i := s_{\min}$ \;}
		$\tau_{i,j}$ is allocated to any available CPUs \;
		The speed of the designated CPU is fixed to $s_i$ \;
	}
	\lElse{No speed reduction can occur. The \edff{k} rule applies; $\tau_{i,j}$ either preempts the lowest priority job currently under execution or is allocated to any available CPU, and the processor speed is fixed to $s_i$.} \;
}
\end{footnotesize}
\caption{Speed-allocation to $\tau_{i,j}$ at time $t$}
\label{ALGO_ALLOC}
\end{algorithm}

Let $s_{i}$ denote the processor speed of the active job $\tau_{i,j}$. This speed $s_i$ is initialized when $\tau_{i,j}$ is released. In a simple version of the \mote\ technique, the execution speed of every released job is initially set to $s_{\offline}$, since we assume that the priorities are assigned by \edff{k} and we proved that the system feasibility is guarantee when it is scheduled by \edff{k} at speed $s_{\offline}$ (Theorem~\ref{Theorem2}). However, we adopt here another initialization step in order to profit from the individual speed of each processor. In this ``optimized'' initialization step, two cases may arise at the arrival of the job $\tau_{i,j}$: 
\begin{enumerate}
\item if $\tau_{i} \in (\tau \setminus \tau^{(k)})$ (the set of the $(k - 1)$ tasks with highest densities), $s_i$ is fixed to $\lambda_i$.
\item if $\tau_{i} \in \tau^{(k)}$, $s_i$ is fixed to $\lambda_k + \frac{\lambda_{\somme}(\tau^{(k+1)})}{m - k + 1}$.
\end{enumerate}
\noindent We proved that all deadlines are met when the system is scheduled while using this rule. Then, when the job $\tau_{i,j}$ is to be allocated to a CPU during the scheduling, we determine the earliest time instant $t_{\next}$ such that $\Pi(\tau_{i,j}, t, t_{\next}) \leq 0$ and if $t_{\next} > t$, one has: 
\begin{equation}
\label{speedreduction}
s_{i} := \min\left\{s_i,\frac{\omega_{i}^{s_{i}}(t) \cdot s_{i}}{\min\left\{D_{i,j},t_{\next}\right\}-t}\right\}
\end{equation}

\noindent We proved also that the system feasibility is not jeopardized by this speed modification.

\subsection{Implementation}
\label{sub:Algorithmic-description}
Before the system starts its execution, our algorithm computes the speed $s_{\offline}$ by determining the optimal value of $k$ thanks to Equation~(\ref{equ:Soffline}) (see Algorithm~\ref{ALGO_INIT}). Then, while the system is running, there is only one kind of situation where the decision to reduce or not the CPU speed for a job $\tau_{i,j}$ is taken: when it is allocated to an available CPU (upon its release, or when it is waiting for an available processor at the head of the ready-Q and a job terminates its execution). A detailed description of the applied procedure at any allocation time is given in Algorithm~\ref{ALGO_ALLOC}. Algorithm~\ref{ALGO_Tnext} shows how to compute $t_{\next}$ with a \emph{linearithmic} (also called \emph{quasilinear}) worst-case computing complexity O($n \cdot \log(n)$), where $n$ is the number of tasks.

It worth noting that the \mote\ step (see Algorithm~\ref{ALGO_ALLOC}) is applied at most once to each job (and only if $i>k$); indeed, a job whose speed has been changed by this step will not be preempted in the future and thus will not be (re-)stored in the \emph{ready-Q} before its end of execution. However, when the speed of a job (with a normal priority) is initialized but not modified by the \mote\ step at its arrival, it can possibly be reduced by the \mote\ step in the future, if the job is at the head of the \emph{ready-Q} and another job completes its execution. Section~\ref{sec:experiments} shows that the \mote\ algorithm indeed \emph{significantly} improves the energy consumption of a real-time sporadic system.

\section{Experiments}
\label{sec:experiments}
\subsection{Introduction}
In our simulations, we have scheduled \emph{periodic constrained-deadline systems} (i.e., $T_i$ is here the exact inter-arrival delay for each task $\tau_i$). The energy consumption of each generated system is computed by simulating the three methods described in this paper during one hyper-period (i.e. the least common multiple of the task periods); indeed, the authors of~\cite{CucuGoossens:07} show that, for the specific case of synchronous periodic task systems, the schedule repeats from the origin with a period equals to the hyper-period. The three methods are: the off-line speed reduction for \edf\ (Equation~(\ref{equ:Sedf})), the off-line speed reduction for \edff{k} (Equation~(\ref{equ:Soffline})) and the \mote\ algorithm (combined with \edff{k}). The energy consumptions generated by these three methods are compared with the consumption by the $S_{\max}$ method (i.e. all jobs are executed at the maximal processors speed $s_{\max}$), while using different processor models. During our simulations, about 5000  constrained-deadline systems were generated and simulated; with the number of tasks $n$ in $[5, 40]$ (with density below 1 and $\lambda_{\somme}(\tau)$ between $1$ and $10$). During each simulation, the ACET of each job was generated using a pseudo-random generator. We made many graphics from our results, but they are omitted here due to space limitation. To ensure that the number $m$ of processors is sufficient to schedule the generated systems at speed $s_{\max}$, $m$ is determined by the following Equation (from~\cite{Goossens2003Priority-driven}): 
\[ m := \min\left\{ n , \left\lceil \frac{\lambda_{\somme}(\tau) - \lambda_{\max}(\tau)}{1 - \lambda_{\max}(\tau)} \right\rceil \right\}\]

\subsection{Processor models}
\label{subsec:realisticmodels}
In our experiments, we used two realistic processor models. These models, noted P1 and P2 in the following, are derived from the processor Crusoe TM5400 from Transmeta and the processor StrongARM SA-1100 from Intel, respectively. In these two processor models, the voltage can only vary in a limited range. Moreover, only a fixed number of functioning frequencies/voltages are available. For that reason, we use the available processor speed immediately above the desired one, if the latter is not available. Note that the use of the two adjacent frequencies to the requested frequency is more efficient from an energy point of view (see, for instance,~\cite{Navet:05}). Table~\ref{table:cpu} (adopted from~\cite{Pouwelse:01} and~\cite{StrongArm}) summarizes the relationship between frequency, voltage, power consumption and the corresponding speed for the Transmeta TM5400 (P1) and the StrongARM SA-1100 (P2). 

\begin{table}[h!]
\centering
\begin{footnotesize}
\begin{tabular}{| c | c | c | c | c |}
\hline
\textbf{CPU} & \textbf{Freq. (MHz)} & \textbf{Volt. (V)} & \textbf{Power (\%)} & \textbf{Speed} \\
\hline
& 700 & 1.65 & 100 & 1\\
\cline{2-5}
& 600 & 1.60 & 80.59 & 0.857 \\
\cline{2-5}
P1 & 500 & 1.50 & 59.03 & 0.714 \\
\cline{2-5}
& 400 & 1.40 & 41.14 & 0.571 \\
\cline{2-5}
& 300 & 1.25 & 24.60 & 0.429 \\
\cline{2-5}
& 200 & 1.10 & 12.70 & 0.286 \\
\hline
\hline
& 206 & 1.50 & 100 & 1 \\
\cline{2-5}
& 195 & 1.42 & 78.9 & 0.947 \\
\cline{2-5}
& 180 & 1.30 & 63.2 & 0.874 \\
\cline{2-5}
& 165 & 1.20 & 50.0 & 0.801 \\
\cline{2-5}
& 150 & 1.15 & 39.9 & 0.728 \\
\cline{2-5}
P2 & 135 & 1.10 & 33.6 & 0.655 \\
\cline{2-5}
& 120 & 1.08 & 33.0 & 0.583 \\
\cline{2-5}
& 105 & 0.95 & 19.8 & 0.510 \\
\cline{2-5}
& 90 & 0.90 & 15.0 & 0.437 \\
\cline{2-5}
& 75 & 0.82 & 11.8 & 0.364 \\
\cline{2-5}
& 60 & 0.80 & 9.44 & 0.291 \\
\hline
\end{tabular}
\end{footnotesize}
\label{table2}
\caption{\label{table:cpu}Processors characteristics.}
\end{table}

Tables~\ref{table4} provides the average consumption profit generated by each method (expressed in percent), compared to the consumption using the $S_{\max}$ method over the entire simulation.

\begin{table}[h!]
\centering
\begin{footnotesize}
\begin{tabular}{| c | c | c |}
\hline
\multicolumn{3}{| c |}{\textbf{results with the StrongARM SA-1100 processor}} \\
\hline
\textbf{Method name} & \textbf{Power saving over $S_{\max}$} & \textbf{Standard deviation} \\
\hline
offline \edf & 4.33 \% & 3.34 \\
\hline
offline \edff{k} & 27.12 \% & 10.24 \\
\hline
\mote & 44.74 \% & 8.82 \\
\hline
\multicolumn{3}{c}{} \\
\hline
\multicolumn{3}{| c |}{\textbf{results with the Crusoe processor}} \\
\hline
\textbf{Method name} & \textbf{Power saving over $S_{\max}$} & \textbf{Standard deviation} \\
\hline
offline \edf & 0.62 \% & 0.76 \\
\hline
offline \edff{k} & 5.91 \% & 4.38 \\
\hline
\mote & 23.3 \% & 7.55 \\
\hline
\end{tabular}
\end{footnotesize}
\caption{Simulation results.}
\label{table4}
\end{table}

\subsection{Observations}
\label{sub:observations}
We observe a large variation in the power saving of our algorithms when they are simulated upon the Crusoe processor and upon the StrongARM SA-1100. This variation is due to the difference in the shape of their consumption function: the consumption function of the StrongARM processor has a higher curvature than the Crusoe processor. That is, a speed reduction in the StrongARM implies a more significant reduction of the system energy consumption. This reduction is therefore \emph{even more significant} when we use the standard dynamic consumption model where the power consumption function is modeled as a constant plus a cubic function (or at least a quadratic function) of the speed~\cite{Zhu:06}. However, our results for this theoretical case are omitted due to the space limitation.

According to~\cite{QuanHu:01}, the Crusoe processor performs a speed transition less than 20~$\mu s$. This time overhead is negligible for most real-time systems, since the order of magnitude of the task characteristics is about few milliseconds. With the Strong ARM SA-1100 processor, Pouwelse et al.~\cite{Pouwelse:01} report that a voltage/speed change can be performed in less than 140~$\mu s$. If this may not be considered as negligible, since we have at most two speed transitions for each job (one initially and one for a \mote\ step), the ``voltage change overheads'' can be incorporated into the worst-case execution requirement. 

\section{Future works}
\label{sec:futureworks}

Currently this work addresses the impact of the proposed scheduling algorithms only on the dynamic power component of the overall microprocessor power dissipation. Proposed methods do not take into account the power dissipated to hold the circuit state and/or power dissipation due to the imperfections of the physical implementation (static power dissipation component). However it is a very well known fact that for integrated circuits manufactured with technologies below 130~$nm$, and especially with current 90~$nm$ and 65~$nm$ technologies, the static power dissipation component becomes very important and comparable to the dynamic power dissipation~\cite{EkekweEC:06}. A significant research effort has been provided, and is still deployed on the static power dissipation reduction techniques. Proposed methods target not only low-level, hardware actions (such as clock gating) but also higher-level (operating system) actions forcing the processor to enter one of the multiple low-power dissipation modes for better trade-off between power saving and wake-up time (see~\cite{Intel:04} as an example). The problem of the increased static power dissipation of the sub-micron technologies is the main motivation for our future work, in which we will extend the existing controllable parameters of our scheduling algorithms (voltage and frequency) with a   processor switch-off parameter.

\section{Conclusion}\label{sec:conclusion}

In this paper, we proposed two approaches which reduce the energy consumption for real-time systems implemented upon multiprocessor platforms. The first one is an adaptation of the first proposal ``Global \edf'', called \edff{k}, which allows a lower computing speed of the processors than \edf. The second proposal (called \mote) is an on-line low-power algorithm which takes into account the ``unused'' CPU times to adjust the processor speeds while the system is running. We show in our experiments that this on-line technique can significantly improve the processors energy consumption (up to $45$\% for the Intel StrongARM SA-1100). Moreover, our \mote\ technique can incorporate the speed/voltage change overheads by simply adding the speed transition time of the processors to the worst-case workload of each task. Our two methods address \emph{sporadic constrained-deadline real-time systems}. This model includes the most popular one: the sporadic and implicit-deadline task systems. The complexity of each decision (at any job allocation-time) is linear in the number of ready jobs in the system. This low-complexity makes the \mote\ strategy a very mighty technique.



\bibliographystyle{acm}
\bibliography{energy}

\begin{thebibliography}{10}

\bibitem{Intel:04}
Intel\textregistered\ pxa27x processor family optimization guide.

\bibitem{Andersson:03}
{\sc Andersson, B.}
\newblock {\em Static-priority scheduling on multiprocessors}.
\newblock PhD thesis, Chalmers Univerosty of Technology, 2003.

\bibitem{AyMeMoMe:04}
{\sc Aydin, R., Melhem, R., Mossé, D., and Mejia-Alvarez, P.}
\newblock Power-aware scheduling for periodic real-time tasks.
\newblock {\em IEEE Transactions on Computers 53}, 5 (2004), 584--600.

\bibitem{BaAn:03}
{\sc Baruah, S., and Anderson, J.}
\newblock Energy-aware implementation of hard-real-time systems upon
  multiprocessor platform.
\newblock In {\em Proceedings of the ISCA 16th International Conference on
  Parallel and Distributed Computing Systems\/} (August 2003), pp.~430--435.

\bibitem{ImprovedScheduleEDF}
{\sc Bertogna, M., Cirinei, M., and Lipari, G.}
\newblock Improved schedulability analysis of {EDF} on multiprocessor
  platforms.
\newblock In {\em ECRTS' 05: Proceedings of the 17th Euromicro Conference on
  Real-Time Systems\/} (2005).

\bibitem{ChenKuoHsu:06}
{\sc Chen, J.-J., Hsu, H.-R., and Kuo, T.-W.}
\newblock Leakage-aware energy-efficient scheduling of real-time tasks in
  multiprocessor systems.
\newblock In {\em 12th IEEE Real-Time and Embedded Technology and Applications
  Symposium\/} (2006), pp.~408--417.

\bibitem{ChenKuo:07}
{\sc Chen, J.-J., and Kuo, T.-W.}
\newblock Energy-efficient scheduling for real-time systems on dynamic voltage
  scaling ({DVS}) platforms.
\newblock In {\em 13th IEEE International Conference on Embedded and Real-Time
  Computing Systems and Applications\/} (August 2007), IEEE Computer Society,
  pp.~28--38.

\bibitem{ChenKuo:06}
{\sc Chen, J.-J., Yang, C.-Y., and Kuo, T.-W.}
\newblock Slack reclamation for real-time task scheduling over dynamic voltage
  scaling multiprocessors.
\newblock In {\em IEEE International Conference on Sensor Networks, Ubiquitous,
  and Trustworthy Computing (SUTC)\/} (Taichung, Taiwan, June 2006).

\bibitem{CucuGoossens:07}
{\sc Cucu, L., and Goossens, J.}
\newblock Feasibility intervals for multiprocessor fixed-priority scheduling of
  arbitrary deadline periodic systems.
\newblock In {\em Design Automation and Test in Europe\/} (2007), IEEE Computer
  Society, pp.~1635--1640.

\bibitem{EkekweEC:06}
{\sc Ekekwe, N., and Etienne-Cummings, R.}
\newblock Power dissipation sources and possible control techniques in ultra
  deep submicron cmos technologies.
\newblock {\em Microelectronics Journal 37}, 9 (September 2006), 851--860.

\bibitem{Ernst97}
{\sc Ernst, R., and Ye, W.}
\newblock Embedded program timing analysis based on path clustering and
  architecture classification.
\newblock In {\em Proceedings of the IEEE/ACM international conference on
  Computer-aided design\/} (California, United States, 1997), IEEE Computer
  Society, pp.~598--604.

\bibitem{Navet:05}
{\sc Gaujal, B., Navet, N., and Walsh, C.}
\newblock Shortest path algorithms for real-time scheduling of fifo tasks with
  optimal energy use.
\newblock In {\em ACM Transactions on Embedded Computing Systems\/} (November
  2005), vol.~4, pp.~907--933.

\bibitem{Goossens2003Priority-driven}
{\sc Goossens, J., Funk, S., and Baruah, S.}
\newblock Priority-driven scheduling of periodic task systems on uniform
  multiprocessors.
\newblock {\em Real Time Systems 25\/} (2003), 187--205.

\bibitem{HoonK:07}
{\sc Kyong~Hoon, K., Rajkumar, B., and Jong, K.}
\newblock Power aware scheduling of bag-of-tasks applications with deadline
  constraints on dvs-enabled clusters.
\newblock In {\em Seventh IEEE International Symposium on Cluster Computing and
  the Grid, 2007. CCGRID 2007\/} (May 2007), pp.~541--548.

\bibitem{Liu:73}
{\sc Liu, C., and Layland, J.}
\newblock Scheduling algorithms for multiprogramming in hard real-time
  environment.
\newblock In {\em Journal of the ACM (JACM)\/} (february 1973), pp.~46--61.

\bibitem{pillai01}
{\sc Pillai, P., and Shin, K.}
\newblock Real-time dynamic voltage scaling for low powered embedded systems.
\newblock {\em Operating Systems Review 35\/} (October 2001), 89--102.

\bibitem{Pouwelse:01}
{\sc Pouwelse, J., Langendoen, K., and Sips, H.}
\newblock Dynamic voltage scaling on a low-power microprocessor.
\newblock In {\em Proceedings of the 7th annual international conference on
  Mobile computing and networking\/} (2001), pp.~251--259.

\bibitem{QuanHu:01}
{\sc Quan, G., and Xiaobo, H.}
\newblock Energy efficient fixed-priority scheduling for real-time systems on
  variable voltage processors.
\newblock In {\em Proceedings of the 38th conference on Design automation\/}
  (2001), pp.~828--833.

\bibitem{ShCh:99}
{\sc Shin, Y., and Choi, K.}
\newblock Power conscious fixed priority scheduling for hard real-time systems.
\newblock In {\em Design Automation Conference\/} (1999), pp.~134--139.

\bibitem{StrongArm}
{\sc Sinha, A., and Chandrakasan, A.~P.}
\newblock Jouletrack: a web based tool for software energy profiling.
\newblock In {\em Proceedings of the 38th conference on Design automation\/}
  (2001), pp.~220--225.

\bibitem{ZhCh:02}
{\sc Zhang, F., and Chanson, S.}
\newblock Processor voltage scheduling for real-time tasks with non-preemptible
  sections.
\newblock In {\em 23th Real-Time Systems Symposium\/} (2002), pp.~235--245.

\bibitem{Zhu:06}
{\sc Zhu, D.}
\newblock Reliability-aware dynamic energy management in dependable embedded
  real-time systems.
\newblock In {\em Proceedings of the 12th IEEE Real-Time and Embedded
  Technology and Applications Symposium, 2006.\/} (April 2006), pp.~397--407.

\end{thebibliography}
\end{document}